\documentclass{article}
\usepackage{amssymb}
\input tcilatex

\begin{document}

\author{Albert Schwarz}
\title{MORITA\ EQUIVALENCE\ AND\ DUALITY}
\date{}
\maketitle

\begin{abstract}
It was shown by Connes, Douglas, Schwarz [1] that one can compactify
M(atrix) theory on noncommutative torus $T_{\theta }$. We prove that
compactifications on Morita equivalent tori are in some sense physically
equivalent. This statement can be considered as a generalization of
non-classical $SL(2,\Bbb{Z})_{N}$-duality conjectured in [1] for
compactifications on two-dimensional noncommutative tori.
\end{abstract}

\section{ Introduction.}

Methods of noncommutative geometry have important applications to M(atrix)
theory. First application of this kind was given in [1], where these methods
were used to describe a new type of toroidal compactifications of M(atrix)
theory. Later such compactifications (compactifications on noncommutative
tori) were studied in numerous papers (see for example [2]).

In present paper we apply to physics the notion of Morita equivalence of
algebras that plays very significant role in noncommutative geometry.
Namely, we prove that the consideration of Morita equivalent noncommutative
tori leads to a kind of duality of compactifications of M(atrix) theory that
generalizes non-classical $SL(2,\Bbb{Z)}_{N}$-duality conjectured in [1] for
two-dimensional case. Our analysis of this duality is based on mathematical
results about Morita equivalence of multidimensional noncommutative tori
obtained in the paper [3] and in this paper.

We work in the framework of IKKT\ M(atrix)\ theory [9]; however by means of
Wick rotation one can obtain corresponding results in BFSS\ M(atrix) model
[8].

Our main results are as follows. We show that compactifications on $n$%
-dimensional noncommutative tori $T_{\theta }$ and $T_{\widehat{\theta }}$
are physically equivalent if antisymmetric matrices $\widehat{\theta }$ and $%
\theta $ are related by the formula $\widehat{\theta }=g\theta =(A\theta
+B)(C\theta +D)^{-1}$ where the block matrix $g$ consisting of $n\times n$
matrices $A,B,C,D$ belongs to the group $SO(n,n|\Bbb{Z})$. ''Vector
bundles'' (projective modules) over noncommutative tori (as well as vector
bundles over commutative tori) can be described by means of integer valued
antisymmetric tensors of even rank, i.e. by even integer elements of
fermionic Fock space. Physically equivalent ''vector bundles'' over $%
T_{\theta }$ and $T_{g\theta }$ are related by linear canonical
transformation in fermionic Fock space; this transformation corresponds to $%
g\in SO(n,n|\Bbb{Z})$. We can consider the space of '' vector bundles'' over
all noncommutative tori; this is a total space of a fibration with a
discrete fibre and a base consisting of all antisymmetric $n\times n$
matrices $\theta $ . It follows from our results that monodromy in this
fibration generates physical equivalence of ''vector bundles''.

In [4] we constructed noncommutative instantons in $\Bbb{R}^{4}$ and
formulated a generalization of Nahm duality for instantons on
four-dimensional noncommutative torus (see [5] for details). It is mentioned
in[4] that Nahm duality is related to Morita equivalence; we are planning to
return to this relation later.

The results of noncommutative geometry that we use are explained in the
paper. The reader willing to learn more about these results can find
necessary information in Connes' book [6] and in papers [7].

Appendix contains some information about $K$-theory and remarks about
relation of $K$-theory to matrix models. It is well known that in $M$-theory
one should consider not only vector bundles, but also more general objects;
it seems that our remarks give an explanation of this fact.

\section{ Basic notions.}

Let us consider a pre- $C^{*}$-algebra $A$ (i.e. an associative algebra over
complex numbers with a norm $\Vert $ $\Vert $ and antilinear involution $*$
obeying the condition $(xy)^{*}$ $=y^{*}x^{*},\Vert xx^{*}\parallel
=\parallel x\parallel ^{2}$).

Let $E$ be a right $A$-module equipped with $A$-valued inner product $<\xi
,\eta >$ that obeys $<\xi ,\eta a>=<\xi ,\eta >a,$ $<\xi ,\eta >^{*}=<\eta
,\xi >,$ $<\xi ,\xi >$ is a positive element of $A$. (Here $\xi ,\eta \in E,$
$a\in A$.) Such a module is called a pre-$C^{*}$-module. $C^{*}$-algebras
and $C^{*}$-modules are defined as complete pre-$C^{*}$-algebras and pre- $%
C^{*}$-modules. We will work with these objects taking completions of pre-$%
C^{*}$-algebras and modules if necessary.

It follows from the definition of inner product that the expression $f(\eta
)=<\xi ,\eta >$ is an $A$ -linear map from $E$ into $A$ .\ We will assume
that every $A$ -linear map from $E$ into $A$ can be represented in this
form. An endomorphism of a $C^{*}$-module $E$ is by definition an $A$-linear
map $T:E\rightarrow E$ having an adjoint map $T^{*}$ (i.e.$T(\xi a)=T(\xi
)\cdot a,$ $<\xi ,T\eta >=<T^{*}\xi ,\eta >,$ where $\xi ,\eta \in E,$ $a\in
A$). One can check that the algebra End$_{A}E$ of endomorphisms of $E$ is a $%
C^{*}$-algebra with respect to the operator norm and involution $%
T\rightarrow T^{*}$.

If $E$ is a finitely generated free module $A^{n}$ (direct sum of $n$ copies
of $A$), then endomorphisms can be identified with $n\times n$ matrices with
entries from $A$.

\emph{In what follows }$A$\emph{\ always stands for a }$C^{*}$\emph{-algebra
having a unit. We will consider only finitely generated projective }$C^{*}$%
\emph{- modules} (i.e. $A$-modules that can be represented as direct
summands in free modules $A^{n}$). Then an endomorphism of an $A$-module $E$
can be defined as any map $T:E\rightarrow E$ obeying $T(\xi a)=T(\xi )a;$
the condition that $T^{*}$ exists is satisfied automatically.

By definition a $C^{*}$-algebra $B$ is (strongly) Morita equivalent to the $%
C^{*}$-algebra $A$ if it is isomorphic to the algebra End$_{A}E$ for some
full module $E$. (One says that $E$ is full if the linear span of the range
of inner product $<\xi ,\eta >$ is dense in $A.$ Then one can prove that
this linear span coincides with $E$ . For important case when $A$ is a
noncommutative torus all modules are full.)

There exists another definition of Morita equivalence that is less
constructive, but much more useful in theoretic considerations. Let us
consider two $C^{*}$-algebras $A,\hat{A}$ and an $(A,\hat{A})$-bimodule $P.$
In other words we assume that the elements of $P$ can be multiplied by
elements of $A$ from the left and by elements of $\hat{A}$ from the right,
that $(a\xi )\hat{a}=a(\xi \hat{a})$ for $a\in A,$ $\hat{a}\in \hat{A},$ $%
\xi \in P$ and that $P$ with these operations can be considered as a left $A$%
-module and right $\hat{A}$-module. As an $A$-module $P$ can be equipped
with $A$-valued inner priduct $<,>_{A}$; as an $\hat{A}$-module it can be
equipped by $\hat{A}$-valued inner product $<,>_{\hat{A}}$; we require that

\begin{equation}
<\xi ,\eta >_{A}\zeta =\xi <\eta ,\zeta >_{\hat{A}}.
\end{equation}

We say that $P$ is an $(A,\hat{A})$-equivalence bimodule if the linear span
of the range of $A$-valued inner product$<,>_{A}$ is dense in $A$ and the
linear span of the range of $\hat{A}$-valued inner product $<,>_{\hat{A}}$
is dense in $\hat{A}.$

One says that $A$ and $\hat{A}$ are Morita equivalent if there exists an $(A,%
\hat{A})$-equivalence bimodule $P.$ To relate this definition to the
definition above we notice that the algebra End$_{\hat{A}}P$ of
endomorphisms of $P$ considered as $\hat{A}$-module is isomorphic to $A$.
Conversely, every full right $A$-module $E$ can be considered as (End$%
_{A}E,A $)- equivalence bimodule.

If $E$ is a right $A$-module and $P$ is an $(A,\hat{A})$ equivalence
bimodule we define a right $\hat{A}$-module $\hat{E}$ by the formula

\begin{center}
\begin{equation}
\hat{E}=E\otimes _{A}P.
\end{equation}
\end{center}

Here the symbol $\otimes _{A}$denotes the tensor product over $A$. (One can
obtain $E\otimes _{A}P$ from the standard tensor product over $\Bbb{C}$ by
means of identification $(\xi a)\otimes \eta \thicksim \xi \otimes (a\eta )$%
. Multiplication by $\hat{a}\in \hat{A}$ in $\hat{E}$ is defined by the
formula $(\xi \otimes \eta )\hat{a}=\xi \otimes (\eta \hat{a})$; this
definition is compatible with the above identification.)

If $P$ is an $(A,\hat{A})$-bimodule then complex conjugate linear space $%
\overline{P}$ can be considered as $(\hat{A},A)$-bimodule. (We use formulas $%
\hat{a}\overline{x}=\overline{x\hat{a}^{*}}$ and $\overline{x}a=\overline{%
a^{*}x}$ to define the multiplication on $\hat{a}\in \hat{A}$ and $a\in A$
in $P$.) In the case when $P$ is an $(A,\hat{A})$ equivalence bimodule, one
can prove that $\overline{P}$ is an $(\hat{A},A)$ equivalence bimodule.
Using $\overline{P}$ we can assign $A$-module to every $\hat{A}$-module; it
is easy to prove that this operation is inverse to the correspondence $%
E\rightarrow \hat{E}$ and therefore the correspondence $E\rightarrow \hat{E}$
can be used to identify equivalence classes of $A$-modules and equivalence
classes of $\hat{A}$-modules. The proof is based on the relation $P\otimes _{%
\hat{A}}\overline{P}=A$ where $A$ is considered as an $(A,A)$-bimodule.

It follows immediately from the definition that to every endomorphism $%
\alpha \in $End$_{A}E$ one can assign an endomorphism $\widehat{\alpha }\in $%
End$_{\hat{A}}\hat{E}$ as a map $\widehat{\alpha }:\hat{E}\rightarrow \hat{E}
$ induced by a map $\alpha \otimes 1:E\otimes _{\Bbb{C}}P\rightarrow
E\otimes _{\Bbb{C}}P.$ (Moreover, one can say that the correspondence $%
E\rightarrow \hat{E}$ can be considered as equivalence of the category $A$%
-modules and the category $\hat{A}$-modules.)

If an algebra $A$ is equipped with a trace Tr$_{A}$ then using inner
products in $(A,\hat{A})$-equivalent bimodule $P$ we can introduce a trace
in $\hat{A}$ by the formula

\begin{equation}
\limfunc{Tr}\ _{\hat{A}}<\xi ,\eta >_{\hat{A}}=\limfunc{Tr}\text{ \ }%
_{A}<\eta ,\xi >_{A}
\end{equation}
where $\xi ,\eta \in P.$ (Recall that linear span of inner products $<\xi
,\eta >$ is dense in $\hat{A}$ by assumption and , moreover, one can prove ,
that it coincides with $\hat{A}$ .) Similarly one can define a trace of
endomorphism $\alpha \in $End$_{A}E$; one can prove that

\begin{equation}
\limfunc{Tr}\widehat{\alpha }=\limfunc{Tr}\alpha .
\end{equation}

Notice that in the case when the trace on $A$ is normalized (i.e. $\limfunc{%
Tr}1=1$) the corresponding trace on End$_{A}E$ is not necessarilly
normalized; its value on unit element is the dimension of $A$-module $E$.
(One can consider this statement as a definition of $\dim E$).

Let us fix a homomorphism of a Lie group $\widetilde{L}$ into the group of
automorphisms of $A.$ It generates a homomorphism of a Lie algebra $L$ into
Lie algebra of derivations (infinitesimal automorphisms) of $A$. The
derivation of $A$ corresponding to $X\in L$ will be denoted by $\delta _{X}$%
; derivations corresponding to the elements of a basis of $L$ will be
denoted by $\delta _{1},...,\delta _{n}.$

Using the Lie algebra $L$ we can define a connection in an $A$-module $E$ as
a set of linear operators $\nabla _{1},...,\nabla _{n}$ acting on $E$ and
obeying the Leibnitz rule:

\begin{equation}
\nabla _{\alpha }(\xi a)=(\nabla _{\alpha }\xi )a+\xi \delta _{\alpha }a.
\end{equation}
Sometimes it is convenient to define connection by means of operators $%
\nabla _{X}$ that depend linearly on $X\in L$ and obey

\begin{equation}
\nabla _{X}(\xi a)=(\nabla _{X}\xi )a+\xi \delta _{X}a
\end{equation}
\emph{We will always consider Hermitian connections} (i.e. the operator $%
\nabla _{\alpha }$ should be antiHermitian with respect to inner product in $%
E$: $<\nabla _{\alpha }\xi ,\eta >+<\xi ,\nabla _{\alpha }\eta >=\delta
_{\alpha }<\xi ,\eta >$).

The curvature $F$ of connection is a $2$-form

\begin{equation}
F_{XY}=[\nabla _{X},\nabla _{Y}]-\nabla _{[X,Y]}.
\end{equation}
This form is defined on the Lie algebra $L$ and takes values in End$_{A}E.$

To work with connections in rigorous way one should consider instead of the
algebra $A$ its dense subset $A^{\infty }$ consisting of elements that are
smooth with respect to the action of the Lie group $\widetilde{L}$ on $A$.
In similar way one should replace an  $A$-module $E$ with its ''smooth
part'' $E^{\infty }.$

\section{ Noncommutative tori.}

By definition, an $n$-dimensional noncommutative torus is an associative
algebra with involution having unitary generators $U_{1},...,U_{n}$ obeying

\begin{equation}
U_{k}U_{j}=e^{2\pi i\theta _{kj}}U_{j}U_{k}
\end{equation}

Notice that different matrices $\theta _{jk}$ can determine isomorphic
algebras; in particular replacing $\theta _{jk}$ by $\theta _{jk}+n_{jk}$
where $n_{jk}\in \Bbb{Z},$ we obtain the same commutation relations and
therefore the same noncommutative torus. In two-dimensional case we consider 
$\theta $ in the notation $T_{\theta }$ as a number (two-dimensional
antisymmetric matrix is determined by one number $\theta =\theta _{12},$
therefore we can identify the matrix with this number).

One can check that the numbers $\theta $ and $\widehat{\theta }$ determine
Morita equivalent two-dimensional tori if 
\begin{equation}
\widehat{\theta }=(A\theta +B)(C\theta +D)^{-1}
\end{equation}
where $A,B,C,D$ are integers and the matrix

\begin{equation}
\left( 
\begin{array}{ll}
A & B \\ 
C & D
\end{array}
\right)
\end{equation}
has determinant 1. Similar statement is correct for multidimensional tori
[3], however in this case one should assume that $A,B,C,D$ are $n\times n$
matrices with integer entries obeying 
\begin{equation}
A^{t}C+C^{t}A=B^{t}D+D^{t}B=0,A^{t}D+C^{t}B=1
\end{equation}
where $^{t}$ denotes transpose.

The noncommutative torus $T_{\theta }$ can be considered as an algebra of
formal expressions

\begin{equation}
\sum C_{X}U_{X}
\end{equation}
where $X$ runs over a lattice $D\subset \Bbb{R}^{n},$

\begin{equation}
U_{X}U_{Y}=e^{\pi i\vartheta _{XY}}U_{X+Y}
\end{equation}
$\vartheta _{XY}$ is a bilinear form on $D$ and $%
U_{X}^{*}=U_{X}^{-1}=U_{-X}. $ (By definition a discrete subgroup $D$ of
vector space $V$ is a lattice if $V/D$ is compact.) It is convenient to
assume that the coefficients $C_{X}$ belong to the Schwartz space $\mathcal{S%
},$ i.e. that they vanish at infinity faster than any power. This assumption
corresponds to consideration of the ''smooth part'' $T_{\theta }^{\infty }$
of the torus $T_{\theta }.$ It is easy to check that

\begin{equation}
U_{X}U_{Y}=e^{2\pi i\theta _{XY}}U_{Y}U_{X}
\end{equation}
where $\theta _{XY}=(\vartheta _{XY}-\vartheta _{YX})/2$ ; therefore
operators $U_{j}=U_{e_{j}}$ where $e_{1},...,e_{n}$ is a lattice basis obey
(8) with $\theta _{jk}=\theta _{e_{j},e_{k}};$ this remark relates the new
description of $T_{\theta }$ with the old one.

If we represent an element of $T_{\theta }$ by means of a complex-valued
function $C_{X}$ on the lattice $D$ then the multiplication in $T_{\theta }$
can be specified by the formula

\begin{equation}
(C*C^{^{\prime }})_{X}=\sum_{Y\in D}e^{\pi i\vartheta
_{Y,X-Y}}C_{Y}C_{X-Y}^{^{\prime }}.
\end{equation}

Instead of $C_{X}$ we can consider its ''Fourier transform''-a function on
the torus $(\Bbb{R}^{n})^{*}/D^{*}$ defined as

\begin{equation}
f(\xi )=\sum_{X\in D}C_{X}e^{2\pi i\xi X}
\end{equation}

Here $D^{*}$ is a lattice dual to $D.$ It consists of such points $\xi \in (%
\Bbb{R}^{n})^{*}$ that $\xi X$ is an integer for all $X\in D.$ We can regard
noncommutative torus $T_{\theta }$ as quantum deformation of commutative
torus $(\Bbb{R}^{n})^{*}/D^{*}.$ The multiplication of functions $%
f,f^{^{\prime }}$ on $(\Bbb{R}^{n})^{*}/D^{*},$ considered as elements of $%
T_{\theta },$ is given by the formula:

\begin{equation}
(f*f^{^{\prime }})(\xi )=(e^{\pi i\theta _{\alpha \beta }\frac{\partial }{%
\partial \xi _{\alpha }}\frac{\partial }{\partial \xi _{\beta }^{^{\prime }}}%
}f(\xi )f^{^{\prime }}(\xi ^{^{\prime }}))_{\xi =\xi ^{^{\prime }}}
\end{equation}

Corresponding commutator can be identified with Moyal bracket.

One can define a trace on the algebra $T_{\theta }$ assigning to an element,
represented by a function $C_{X}$ on the lattice, the value of this function
at $X=0$ . The trace, defined by means of this construction is unique ( up
to a factor).

To construct a module over $T_{\theta }$ we should find a Hilbert space $%
\mathcal{H}$ and unitary operators $U_{X}$ on $\mathcal{H}$ obeying (14).
Let us work with right modules and use the notation $\xi U_{X}$ where $\xi
\in \mathcal{H}.$ To introduce a structure of $C^{*}$-module in $\mathcal{H}$
we should define a $T_{\theta }$-valued inner product \TEXTsymbol{<} , 
\TEXTsymbol{>} starting with $\Bbb{C}$-valued inner product ( \ , ). This
can be done by means of the formula

\begin{equation}
<\xi ,\eta >=\sum_{X\in D}(\xi ,\eta U_{-X})U_{X}.
\end{equation}

Under certain assumptions about convergence of the series in (18) it is easy
to check that the inner product \TEXTsymbol{<} , \TEXTsymbol{>} satisfies
the conditions in the definition of $C^{*}$-module. It is more difficult to
formulate easily verifiable conditions of projectivity of the module. ( The
most convenient way to prove projectivity is to check that the unit operator
can be represented as an endomorphism of finite rank, i.e. as an operator
defined by the formula $x\rightarrow \sum <\xi _{i},x>\eta _{i}$ ).

The situation in the case of left modules is the same. (Every left $%
T_{\theta }$-module can be considered as a right $T_{-\theta }$-module.)

Examples of $T_{\theta }$-modules can be obtained in the following way. Let
us define operators $U_{\gamma \widetilde{,\gamma }}$ in the space $E=L^{2}(%
\Bbb{R}^{n})$ by the formula: 
\begin{equation}
(U_{\gamma \widetilde{,\gamma }}f)(X)=e^{2\pi i\widetilde{\gamma }%
X}f(X+\gamma ).
\end{equation}

Here $\gamma \in \Bbb{R}^{n},\widetilde{\gamma }\in (\Bbb{R}^{n})^{*}.$ It
easy to check that

\begin{equation}
U_{\gamma \widetilde{,\gamma }}U_{\lambda ,\widetilde{\lambda }}=e^{2\pi i%
\widetilde{\gamma }\lambda }U_{\gamma +\lambda ,\widetilde{\gamma }+%
\widetilde{\lambda }}.
\end{equation}

Using this relation one can construct a left $T_{\theta }$-module $E_{\Gamma
}$ by means of operators $U_{\gamma ,\widetilde{\gamma }}$ where $(\gamma ,%
\widetilde{\gamma })\in \Gamma \subset V=\Bbb{R}^{n}+(\Bbb{R}^{n})^{*}$ and $%
\Gamma $ is a lattice in $V$. (The requirement that $\Gamma $ be a lattice
is necessary to prove that $E_{\Gamma }$ is a projective module.) It follows
from (20) that the operators $U_{\gamma ,\widetilde{\gamma }},$ $U_{\lambda ,%
\widetilde{\lambda }}$, where $(\gamma ,\widetilde{\gamma }),(\lambda ,%
\widetilde{\lambda })\in \Gamma $ obey (14) with $\theta _{(\gamma ,%
\widetilde{\gamma }),(\lambda ,\widetilde{\lambda })}=\widetilde{\gamma }%
\lambda -\widetilde{\lambda }\gamma ,$ and therefore specify a module over
noncommutative torus. ( More precisely, to define $T_{\theta }$ -valued
inner product we should restrict the class of functions taking only
functions from the Schwartz class $S(\Bbb{R}^{n})$.)

Let us define the lattice $\Gamma ^{*}$ as a group of all $(\mu ,\widetilde{%
\mu })\in \Bbb{R}^{n}+(\Bbb{R}^{n})^{*}$obeying the condition

\begin{equation}
\widetilde{\mu }\gamma -\widetilde{\gamma }\mu \in \Bbb{Z}.
\end{equation}

It is easy to check that every operator $U_{\mu ,\widetilde{\mu }}$ where $%
(\mu ,\widetilde{\mu })\in \Gamma ^{*}$ commutes with all operators $%
U_{\gamma ,\widetilde{\gamma }}$ where $(\gamma ,\widetilde{\gamma })\in
\Gamma .$ One can prove that every endomorphism of $E_{\Gamma }$ belongs to
the closure \ ( for the $C^{*}$ -norm ) of the linear span of $U_{\mu ,%
\widetilde{\mu }}$ where $(\mu ,\widetilde{\mu })\in \Gamma ^{*}.$ This
means that End$_{T_{\theta }}E_{\Gamma }$ can be identified with the torus $%
T_{\widehat{\theta }}$, where $\widehat{\theta }_{(\mu ,\widetilde{\mu }%
)\left( \nu ,\widetilde{\nu }\right) }=-\widetilde{\mu }\nu +\widetilde{\nu }%
\mu $ is a bilinear form on $\Gamma ^{*}.$ The torus $T_{\widehat{\theta }}$
is Morita equivalent to $T_{\theta }.$ The space $S(\Bbb{R}^{n})$ considered
as a left $T_{\theta }$-mobile $E_{\Gamma }$ and a right $T_{\widehat{\theta 
}}$-mobile $E_{\Gamma ^{*}}$ is a $(T_{\theta },T_{\widehat{\theta }})$%
-equivalence bimodule; we will denote this bimodule by $P_{\Gamma ,\Gamma
^{*}}.$

The above construction can be generalized in the following way. Let us
consider instead of $\Bbb{R}^{n}$ any abelian group $G$ that can be
represented as a direct sum of $\Bbb{R}^{n}$ and finitely generated abelian
group $G^{^{\prime }}$ (in other words $G^{^{\prime }}$ is a direct sum of
several copies of $\Bbb{Z}$ and finite cyclic groups $\Bbb{Z}_{m}$). Then we
can consider operators $(U_{\gamma ,\widetilde{\gamma }}f)(x)=e^{2\pi i%
\widetilde{\gamma }(x)}f(x+\gamma )$ where $x\in G,\gamma \in G,\widetilde{%
\gamma }\in G^{*}.$ Here $G^{*}$ stand for the group of characters of $G$
(i.e. a group\thinspace of continuous homomorphisms of $G$ into the group $K=%
\Bbb{R}/\Bbb{Z}$).

Every lattice $\Gamma \subset G\times G^{*}$ determines a module over
noncommutative torus. Endomorphisms of this module can be described by means
of dual lattice $\Gamma ^{*},$ that consists of elements $(\mu ,\widetilde{%
\mu })\in G\times G^{*}$ obeying the condition

\begin{equation}
\widetilde{\mu }\gamma -\widetilde{\gamma }\mu =0
\end{equation}
for all $(\gamma ,\widetilde{\gamma })\in \Gamma .$ (Notice that $\widetilde{%
\mu }\gamma =\widetilde{\mu }(\gamma )$ and $\widetilde{\gamma }\mu =%
\widetilde{\gamma }(\mu )$ are considered as elements of $K=\Bbb{R}/\Bbb{Z}$%
).

In particular, for two-dimensional noncommutative torus $T_{\theta }$ we can
take $G=\Bbb{R\times Z}_{q}$ and a lattice $\Gamma _{p,q}$ spanned by
elements $(\gamma _{1},\widetilde{\gamma }_{1}),(\gamma _{2},\widetilde{%
\gamma }_{2})$ where $\gamma _{1}=(0,-p),\gamma _{2}=(1,0),\widetilde{\gamma 
}_{1}=(p/q-\theta ,0),\widetilde{\gamma }_{2}=(0,-1).$ (We consider $\Bbb{Z}%
_{q}$ as $\Bbb{Z}/qZ$ and identify $G^{*}$ with $\Bbb{R}\times \Bbb{Z}_{q}$%
.) Corresponding $T_{\theta }$-module will be denoted by $E_{p,q}$. (Here $%
p\in \Bbb{Z},q\in \Bbb{Z}$.)

Let us consider as an example the case when the lattice $\Gamma \subset \Bbb{%
R}^{2}$ is spanned by the vectors $(1,0),(0,\theta +n)$ where $n\in \Bbb{Z}%
,\theta \in \Bbb{R}.$ The lattice $\Gamma ^{*}$ is spanned by vectors $(1,0)$
and $(0,(\theta +n)^{-1})$. We obtain that two-dimensional tori $T_{\theta }$
and $T_{(\theta +n)^{-1}}$ are Morita equivalent. (Recall that the tori $%
T_{\theta }$ and $T_{\theta +n}$ coincide.) We will analyze the relation
between $T_{\theta }$-modules and $T_{(\theta +n)^{-1}}$-modules that stems
from Morita equivalence between $T_{\theta }$ and $T_{(\theta +n)^{-1}}.$

To every pair $(p,q)$ of integers we assigned a $T_{\theta }$-module $%
E_{p,q} $; if $\theta $ is irrational we obtain a one-to one correspondence
between pairs $(p,q)\in \Bbb{Z}^{2}$ obeying $p-q\theta >0$ and (classes of
projective ) modules. It follows from our general results (see Sec.5) that $%
T_{(\theta +n)^{-1}}$-module related to $T_{\theta }$-module $E_{p,q}$
corresponds to the pair $(-q,-(p+qn)).$

We will give now a direct proof of this result for the case of the module $%
E_{p,1}.$ Such a module can be described as a right module $E_{\Gamma }$
corresponding to a lattice $\Gamma _{p,1}$ spanned by $(1,0),(0,\theta +p).$
The tensor product $E_{\Gamma _{p,1}}\otimes _{\Bbb{C}}P_{\Gamma ,\Gamma
^{*}}$ can be regarded as a space of functions $f(x,y)$ depending on two
variables should identify $f(x+1,y)$ with $f(x,y+1)$ and $x\in \Bbb{R},y\in 
\Bbb{R}$. To obtain $\widehat{E}=E_{\Gamma _{p,1}}\otimes _{T_{\theta
}}P_{\Gamma ,\Gamma ^{*}}$ we$e^{-2\pi i(\theta +p)x}f(x,y)$ with $e^{2\pi
i(\theta +n)y}f(x,y).$ It is easier to describe the dual space $\widehat{E}%
^{*}$ that consists of generalized functions $g(x,y)$ obeying

\begin{equation}
g(x+1,y)=g(x,y+1)
\end{equation}

\begin{equation}
e^{-2\pi i(\theta +p)x}g(x,y)=e^{2\pi i(\theta +n)y}g(x,y)
\end{equation}

One can check that an element of $\widehat{E}^{*}$ (a solution to (23),(24))
can be represented in the form

\begin{equation}
g(x,y)=\sum_{k}\delta (\theta (x+y)+px+ny-k)g_{k}(x+y)
\end{equation}
where $g_{k}(x)=g_{k+(p+n)}(x).$ In other words we can identify elements of $%
(\widehat{E})^{*}$ with functions on $\Bbb{R}\times \Bbb{Z}_{p+n}.$ By
definition the action of $T_{(\theta +n)^{-1}}$ on $\widehat{E}$ is induced
by \ the action of $T_{(\theta +n)^{-1}}$ on $P_{\Gamma ,\Gamma ^{*}}.$ This
means that we should consider operators on $E\otimes _{\Bbb{C}}P_{\Gamma
,\Gamma ^{*}}$ that transform $f(x,y)$ into $f(x,y+1)$ and $e^{2\pi i(\theta
+n)^{-1}y}f(x,y)$ respectively. These operators induce operators on $%
\widehat{E}$ and on $(\widehat{E})^{*}.$ It is easy to describe explicitly
the induced operators on $(\widehat{E})^{*}$ . It follows from this
description that $T_{(\theta +n)^{-1}}$-module $\widehat{E}$ corresponds to
a pair $(-1,-(p+n)).$

\section{ Compactifications of M(atrix) theory.}

Let us consider the case when the $C^{*}$-algebra $A$ has a trace Tr and $L$
is a ten-dimensional commutative Lie algebra, equipped with inner product (
, ).

Let us fix an orthonormal basis in $L$ with respect to this inner product.
Then for every $A$-module $E$ we can define a functional $I$ by the formula 
\begin{equation}
I=\sum_{\alpha ,\beta }\limfunc{Tr}F_{\alpha \beta }^{2}+2\sum \limfunc{Tr}%
\Psi ^{i}\Gamma _{ij}^{\alpha }[\nabla _{\alpha },\Psi ^{j}]
\end{equation}

Here $F_{\alpha \beta }$ are components of the curvature tensor of
connection $\nabla _{\alpha }$ with respect to the orthonormal basis in $L$
and $\Psi ^{i}$, $i=1,...,16$ are elements of $\Pi $End$_{A}E$, where $\Pi $
stands for parity reversing. The symbol $\Gamma _{ij}^{\alpha }$ denotes
ten-dimensional Dirac matrices.

We consider $I$ as a functional on Conn$\times (\Pi $\ End$_{A}E\otimes S)$
where Conn stands for the space of all Hermitian connections on $E$ and $S$
is the space of Weyl spinors.

In the case when $A$ is the algebra of complex numbers, $E=\Bbb{C}^{N}$ is
an $N$-dimensional vector space (free module) and the algebra $L$ acts
trivially on $\Bbb{C}$ one can identify (26) with the action functional of
IKKT matrix model.

In the case when $A$ is a noncommutative torus one can consider (26) as an
action functional of toroidal compactification of IKKT\ M(atrix) theory;
this fact follows from the results of [1]. To verify this statement we
notice that the relations (8) remain correct if we replace $U_{k}$ with $%
\widetilde{U}_{k}=\lambda _{k}U_{k}$ where $\mid \lambda _{k}\mid =1$. This
means that we can consider an $n$-dimensional Lie algebra $L_{\theta }$ of
derivations of $T_{\theta }$; the generators of $L_{\theta }$ act by the
formula $\delta _{j}U_{k}=iU_{k}$ if $j=k$, $\delta _{j}U_{k}=0$ if $j\neq
k. $ An action of ten-dimensional commutative algebra $L$ on $T_{\theta }$
is defined by means of an arbitrary surjective linear map of $L=R^{10}$ onto 
$L_{\theta }=R^{n}$. Corresponding connections can be identified with the
solutions to Eqn (3.14) of [1].

One can modify the functional (26) replacing $F_{\alpha \beta }$with $%
F_{\alpha \beta }+\varphi _{\alpha \beta }\cdot 1$ where $\varphi _{\alpha
\beta }$ is an antisymmetric $2$-form on $L$. We obtain a functional 
\begin{equation}
J=\sum_{\alpha ,\beta }\limfunc{Tr}(F_{\alpha \beta }+\varphi _{\alpha \beta
}\cdot 1)^{2}+2\sum \limfunc{Tr}\Psi ^{i}\Gamma _{ij}^{\alpha }[\nabla
_{\alpha },\Psi ^{j}]
\end{equation}
that also was considered in [1].

The functional (27) depends on the following data (in the case $A=T_{\theta
}).$

a) symmetric bilinear form $g_{\alpha \beta }$ on $L=R^{10}$ that determines
an inner product on $L$ (we work in orthonormal basis, therefore $g_{\alpha
\beta }=\delta _{\alpha \beta }$);

b) antisymmetric bilinear form $\varphi _{\alpha \beta }$ on $L=R^{10}$;

c) antisymmetric bilinear form $\theta _{\alpha \beta }$ on $R^{n}$,
specifying the algebra $T_{\theta }$;

d) surjective map $\delta :L=R^{10}\rightarrow L_{\theta }=R^{n}$;

e) $T_{\theta }$-module $E.$

We will show that using Morita equivalence we can transform this set of data
into another set of data giving a physically equivalent functional (27).

Our starting point will be an $(A,\hat{A})$-equivalence bimodule $P$, i.e. a
bimodule that gives Morita equivalence between $A=T_{\theta }$ and $\hat{A}%
=T_{\widehat{\theta }}$. (Our consideration is valid not only for tori,
therefore we prefer to use more general notations.)We will assume that there
exists a constant curvature connection $\nabla _{X}^{P}$ on $P$, considered
as a left $A$-module. In other words, we require that

\begin{equation}
\nabla _{X}^{P}(a\xi )=a\nabla _{X}^{P}\xi +(\delta _{X}a)\cdot \xi ,
\end{equation}

\begin{equation}
\lbrack \nabla _{X}^{P},\nabla _{Y}^{P}]=\sigma _{XY}\cdot 1
\end{equation}
Here $\xi \in P,a\in A,X\in L,\delta _{X}$ stands for the action of $L$ on $%
A $, $\sigma _{XY}$ is an antisymmetric bilinear form on $L$. We impose a
condition

\begin{equation}
\nabla _{X}^{P}(\xi \hat{a})=(\nabla _{X}^{P}\xi )\hat{a}+\xi \widehat{%
\delta }_{X}\hat{a},
\end{equation}
where $\xi \in P,\hat{a}\in \hat{A}$. This means that $\nabla _{X}^{P}$ is
also a connection in the right $\hat{A}$-module $P$ for an appropriate
definition of an action $\widehat{\delta }_{X}$ of $L$ on $\hat{A}.$
(Talking about an action of commutative Lie algebra $L$ on $A=T_{\theta }$
and on $\widehat{A}=T_{\widehat{\theta }}$ we will have in mind
homomorphisms $\delta $ and $\widehat{\delta }$ from $L$ into $L_{\theta }$
and $L_{\widehat{\theta }}$ respectively; we assume that these homomorphisms
are surjective.)

We will say that the $(A,\hat{A})$-module $P$ satisfying the conditions
above determines complete Morita equivalence between $A$ and $\hat{A}$ . It
is easy to check that the bimodule $P_{\Gamma ,\Gamma ^{*}}$ constructed
above determines complete Morita equivalence between $T_{\theta }$ and $T_{%
\widehat{\theta }}$.

Recall, that using the $(A,\hat{A})$-module $P$ we can construct a
one-to-one correspondence between right $A$-modules and right $\hat{A}$%
-modules by the formula $\hat{E}=E\otimes _{A}P$.

If $\nabla _{X}$ is a connection in the right $A$-module $E$ (i.e. it obeys
(6)) we can consider $\widehat{\nabla }_{X}=\nabla _{X}\otimes 1+1\otimes
\nabla _{X}^{P}$ as on operator on $\hat{E}$. (One can obtain $\hat{E}%
=E\otimes _{A}P$ from $E\otimes _{\Bbb{C}}P$ by means of identification $\xi
a\otimes \eta \sim \xi \otimes a\eta $. It is clear that $\widehat{\nabla }%
_{X}$ acts on $E\otimes _{\Bbb{C}}P$; to verify that $\widehat{\nabla }_{X}$
acts on $\hat{E}$ we should check the compatibility with this
identification.)

Using (30)\ we check, that $\widehat{\nabla }_{X\text{ }}$ is a connection
in the right $\hat{A}$-module $\hat{E}$. Let us calculate its curvature $%
F_{XY}^{\widehat{\nabla }}=F_{XY}^{\wedge }$. Notice that the curvature $%
F_{XY}=F_{XY}^{\nabla }$ of the connection $\nabla _{X}$ for fixed $X,Y\in L$
can be considered as an endomorphism of $E$; therefore we can construct the
corresponding endomorphism $\widehat{F}_{XY}$ of $\hat{E}.$ It is easy to
see that

\begin{equation}
F_{XY}^{\wedge }=\widehat{F}_{XY}+\sigma _{XY}\cdot 1
\end{equation}
This follows from the remark that considering $\widehat{\nabla }_{X}$ as an
operator on $E\otimes _{\Bbb{C}}P$ we obtain

\begin{equation}
\lbrack \widehat{\nabla }_{X},\widehat{\nabla }_{Y}]=F_{XY}+\sigma
_{XY}\cdot 1
\end{equation}

Both terms in the RHS of (32)\ induce operators on $\hat{E}=E\otimes _{A}P$;
we immediately obtain (31) from (32) and the definition of $F_{XY}^{\wedge
}. $ The formula (31) permits us to derive some important statements. We see
first of all that in the case when the connection $\nabla _{X}$ has constant
curvature the corresponding connection $\widehat{\nabla }_{X}$ also has
constant curvature. (It equals to $(f_{XY}+\sigma _{XY})\cdot 1$, where $%
f_{XY}$ stands for the curvature of $\nabla _{X}$.) In other words, if $%
\nabla _{X}$ determines a BPS\ state having maximal supersymmetry, then $%
\widehat{\nabla }_{X}$ has the same property. Analogous statement is true
for BPS\ states having less supersymmetries. Recall that $\nabla _{X}$
determines a BPS state if one can find spinors $\varepsilon ,\varepsilon
^{^{\prime }}$ obeying

\begin{equation}
\varepsilon \Gamma ^{\alpha \beta }F_{\alpha \beta }+\varepsilon ^{^{\prime
}}\cdot 1=0
\end{equation}
It follows from (31)\ that then

\begin{equation}
\varepsilon \Gamma ^{\alpha \beta }F_{\alpha \beta }^{\wedge }+\varepsilon
^{^{\prime \prime }}\cdot 1=0
\end{equation}
where $\varepsilon ^{^{\prime \prime }}=\varepsilon ^{^{\prime
}}-\varepsilon \Gamma ^{\alpha \beta }\sigma _{\alpha \beta }$. We see that $%
\widehat{\nabla }_{X}$ determines a BPS\ state having the same number of
supersymmetries as $\nabla _{X}.$

Now we can represent the action functional $J$ defined on the space Conn $%
\times ($End$_{A}E\otimes S)$ as an action functional $\widehat{J}$ defined
on the space $\widehat{\limfunc{Conn}}\times ($End$_{\hat{A}}\widehat{E}%
\otimes S),$ where $\widehat{\limfunc{Conn}}$ stands for the space of
connections on $\hat{A}$-module $\hat{E}$, that are defined by means of the
action $\widehat{\delta }_{X}$ of $L$ on $\hat{A}$. Expressing $J$ in terms
of connections $\widehat{\nabla }_{X}$ and endomorphisms $\widehat{\Psi }%
^{i} $ corresponding to $\nabla _{X}$ and $\Psi ^{i}$ we obtain

\begin{equation}
\widehat{J}=\sum_{\alpha ,\beta }\limfunc{Tr}(\widehat{F}_{\alpha \beta
}+(\varphi _{\alpha \beta }-\sigma _{\alpha \beta })\cdot 1)^{2}+\sum 
\limfunc{Tr}\widehat{\Psi }^{i}\Gamma _{ij}^{\alpha }[\widehat{\nabla }%
_{\alpha },\widehat{\Psi }^{j}].
\end{equation}
To be sure that the theory based on the consideration of the action
functional $\widehat{J}$ is equivalent to the theory based on $J$ we should
check that the correspondence Conn$\rightarrow \widehat{\limfunc{Conn}}$ we
constructed is bijective. The proof is based on the construction of the
inverse map $\widehat{\limfunc{Conn}}\rightarrow $Conn. Recall that to prove
bijectivity of map $E\rightarrow \hat{E}=E\otimes _{A}P$ we used an $(\hat{A}%
,A)$-equivalence bimodule $\overline{P\text{.}}$ Here we use the same
bimodule equipped with a connection $\overline{\nabla }_{\alpha }$ defined
by the formula 
\[
\overline{\nabla }_{\alpha }(\overline{\omega })=\overline{(\nabla _{\alpha
}\omega ).} 
\]
The fact, that $E=\hat{E}\otimes _{\hat{A}}\overline{P}$ follows from the
remark that $P\otimes _{\hat{A}}\overline{P}=A$ as $(A,A)$-bimodule. To
prove similar fact for connections we notice that the operator $\nabla
_{\alpha }\otimes 1+1\otimes \overline{\nabla }_{\alpha }$ on $P\otimes _{%
\Bbb{C}}\overline{P}$ induces operator $\delta _{\alpha }$ on $A=P\otimes _{%
\hat{A}}\overline{P}$. To check this we notice that the natural map of $%
P\otimes _{\Bbb{C}}\overline{P}$ onto $P\otimes _{\hat{A}}\overline{P}=A$
can be identified with inner product $<\cdot ,\cdot >_{A}$ in $P$. Then the
statement we need follows immediately from the assumption that $\nabla
_{\alpha }$ is a Hermitian connection.

\section{ Detailed analysis of toroidal compactifications.}

Let us give a detailed analysis of the above constructions for the case when 
$A$ is a noncommutative torus $T_{\theta }$. It is shown in [3] that $T_{%
\widehat{\theta }}$ is Morita equivalent to $T_{\theta }$ if $\widehat{%
\theta }=g\theta $ where $g\in SO(n,n|\Bbb{Z})$, i.e. $g$ is a matrix of the
form (10) having integral entries and unit determinant and obeying (11).
(The definition of the action of $SO(n,n|\Bbb{Z})$ on a dense subset of the
space of antisymmetric matrices is given by the formula(9) .) It is easy to
check that the proof of Morita equivalence in [3] gives complete Morita
equivalence $T_{\theta }\sim T_{g\theta }$ in the sense of present paper
(the $(T_{\theta },T_{g\theta })$-equivalence bimodule constructed in [3]
can be represented as a bimodule $P_{\Gamma ,\Gamma ^{*}}$ for appropriate
choice of lattices $\Gamma ,\Gamma ^{*}\subset G\times G^{*}$ and therefore
it can be equipped with a connection obeying the conditions we imposed). We
will prove that, conversely, if the noncommutative $T_{\widehat{\theta }}$
is completely Morita equivalent to the torus $T_{\theta }$ then $\widehat{%
\theta }$ and $\theta $ belong to the same orbit of the group $SO(n,n|\Bbb{Z}%
)$. In other words, we will see that $T_{\widehat{\theta }}$ is completely
Morita equivalent to $T_{\theta }$ if and only if $\widehat{\theta }=g\theta 
$ for some $g\in SO(n,n|\Bbb{Z})$.

We will give also an explicit description of the correspondence between
projective modules over $T_{\theta }$ and projective modules over $T_{%
\widehat{\theta }}$.

We will start with formulation of some well known results. Notice that the
Lie algebra $L_{\theta }$ of derivations of $T_{\theta }$ can be considered
as a Lie algebra of a Lie group $\widetilde{L}_{\theta }$ of automorphisms
of $T_{\theta }$. The Lie group $\widetilde{L}_{\theta }$ is a torus that
can be identified with $L_{\theta }/D$, where $D=\Bbb{Z}^{n}$ is a lattice
in $L_{\theta }$. One can consider an element of $\widetilde{L}_{\theta }$
as a row $(\lambda _{1},...,\lambda _{n})$ where $|\lambda _{i}|=1$; such a
row acts on $T_{\theta }$ transforming $U_{j}$ into $\lambda _{j}U_{j}$,
where $U_{j}$ are generators of $T_{\theta }.$ Then $D$ can be regarded as a
lattice generated by $e_{j}=2\pi \delta _{j}\in L_{\theta }$ where $\delta
_{j}U_{k}=iU_{k}$ for $j=k,\delta _{j}U_{k}=0$ for $j\neq k$. Let us denote
by $\mathcal{F}$ the Grassmann algebra $\Lambda ($ $L_{\theta })$. An
element of $\mathcal{F}$ can be considered as a formal linear combination

\begin{equation}
\omega =\sum_{k}\sum_{j_{1}...j_{k}}\omega
^{j_{1}...j_{k}}e_{j_{1}}...e_{j_{k}}
\end{equation}
where $\omega ^{j_{1}...j_{k}}\in \Bbb{C}$ is antisymmetric with respect to $%
j_{1},...j_{k}$. Requiring that $\omega ^{j_{1}...j_{k}}$ in (36) be
integers we single out a subring $\Lambda (D)\subset \Lambda (L_{\theta })$.
Similarly, using the lattice $D^{*}\subset L^{*}$ that is dual the lattice $%
D\subset L$ we define the ''integral part'' $\Lambda (D^{*})$ of $\mathcal{F}%
^{*}=\Lambda (L^{*})$. One can identify the group $K_{0}(T_{\theta })$ with
even part $\Lambda ^{ev}(D^{*})$ of $\Lambda (L^{*})$ (the ring $\Lambda
(D^{*})$ inherits $\Bbb{Z}_{2}$-grading from $\Lambda (L_{\theta }^{*})$; we
define its even part with respect to this grading). Recall that to define $%
K_{0}(A)$ one should apply the Grothendieck construction to the semigroup of
equivalence classes of projective $A$-modules with respect to direct sum.
(In other words we consider formal differences $E_{1}-E_{2}$ where $%
E_{1},E_{2}$ are projective $A$-modules and identify $(E_{1}-E_{2})+E_{2}$
with $E_{1}$; see Appendix for more information about $K_{0}$.) In
particular, every projective $A$-module $E$ determines an element $\mu
(E)\in K_{0}(A)$; the set $K_{0}^{+}(A)$ of elements of $K_{0}(A)$ that can
be obtained this way is by definition the positive cone in $K_{0}(A)$. In
the case $A=T_{\theta }$ and $\theta $ is irrational (i.e. at least one
entry of the matrix $\theta $ is irrational) one can prove that two
non-equivalent projective $A$-modules determine different elements of $%
K_{0}(E)$; this permits us to identify a projective $A$-module with its
representative $\mu =\mu (E)\in K_{0}^{+}(A)$.

The Chern character ch $E$ of a projective $A$-module $E$ is defined by the
formula

\begin{equation}
\limfunc{ch}E=\widehat{\tau }(e^{F/2\pi i})=\sum_{n=0}\frac{1}{n!}\widehat{%
\tau }(F^{n})\cdot \frac{1}{(2\pi i)^{n}}
\end{equation}
where $F$ is a curvature of an arbitrary connection on $E$. Recall that to
define a connection we fixed a homomorphism $\delta $ of a Lie algebra $L$
into the Lie algebra of derivations of $A$. Then the curvature $F$ can be
considered as a two-form on $L$ taking values in the algebra $\widehat{A}=$%
End$_{A}E$ and $F^{n}$ is a $2n$-form taking values in \thinspace $\widehat{A%
}\otimes ...\otimes \widehat{A}$. We assume that the algebra $A$ is equipped
with a trace $\tau $; as we mentioned already then we can define a trace $%
\widehat{\tau }$ on $\widehat{A}$ and therefore on $\widehat{A}\otimes
...\otimes \widehat{A}$. The formula (37) determines the Chern character ch$%
(E)$ as an inhomogeneous form on $L$ (i.e. as an element of the Grassmann
algebra $\mathcal{F}^{*}=\Lambda (L^{*})$ that is dual to $\mathcal{F}%
=\Lambda (L)$. (Notice that the definition of connection that we use is a
particular case of more general definition that we do not need. In this
general definition the Chern character takes values in cyclic homology of $A$%
.)

If $A$ is a noncommutative torus $T_{\theta }$ one can consider the matrix $%
\theta $ as a $2$-form on $L=L_{\theta }$, i.e. as an element of $\theta \in
\Lambda (L)$. Then one can prove the formula : 
\begin{equation}
\limfunc{ch}E=e^{\theta }\lrcorner \mu (E),
\end{equation}
where $\lrcorner $ stands for the operation of contraction. It is convenient
to consider elements $\mathcal{F}^{*}=\Lambda (L^{*})$ as functions of
anticommuting variables $\alpha ^{1},...,\alpha ^{n}$. Then one can write
the formula (38)$\;$in the form

\begin{equation}
chE=e^{\frac{1}{2}b_{k}\theta ^{kj}b_{j}}\mu (E)
\end{equation}
where $b_{j}$ stands for the derivative with respect anticommuting variable $%
\alpha ^{j}$. One can consider $\mathcal{F}^{*}$ also as a fermionic Fock
space where $b_{1},...,b_{n}$ play the role of annihilation operators, and
operators $a^{i}$defined by means of multiplication by $\alpha ^{i}$ play
the role of creation operators. (The operators $a^{i},b_{i}$ satisfy
canonical anticommutation relations $[b_{k},a^{j}]_{+}=\delta
_{k}^{j},[a^{j},a^{k}]_{+}=[b_{j},b_{k}]_{+}=0$

Let us fix complete Morita equivalence between $A=T_{\theta }$ and $\widehat{%
A}=T_{\widehat{\theta }}$. We will study the relation between $\mu =\mu (E)$
and $\widehat{\mu }=\mu (\widehat{E})$ where $\widehat{A}$-module $\widehat{E%
}$ corresponds to the $A$-module $E$. Recall, that for every connection $%
\nabla $ in $E$ we constructed a connection $\widehat{\nabla }$ in $\widehat{%
E}$ and that the curvatures of $\nabla $ and $\widehat{\nabla }$ are related
by the formula (31). It follows from (37),(31) that 
\begin{equation}
\limfunc{ch}E=\widehat{\tau }\exp (\frac{1}{2\pi i}\alpha ^{k}F_{kj}^{\nabla
}\alpha ^{j})
\end{equation}

\begin{equation}
\limfunc{ch}\widehat{E}=\tau \exp (\frac{1}{2\pi i}\alpha ^{k}F_{kj}^{%
\widehat{\nabla }}\alpha ^{l})=e^{\frac{1}{2\pi i}\alpha ^{k}\varphi
_{kj}\alpha ^{l}}\limfunc{ch}E
\end{equation}
where we identified $L_{\theta }$ and $L_{\widehat{\theta }}$ with $\Bbb{R}%
^{n}$ in the way that was used in (31). Applying (39),(40),(41) we relate $%
\mu $ and $\widehat{\mu }$ in the following way:

\begin{equation}
\widehat{\mu }=V_{1}V_{2}V_{3}V_{4}\mu
\end{equation}
where 
\begin{equation}
V_{1}f=\exp (-\frac{1}{2}b_{k}\widehat{\theta }^{kj}b_{j})f
\end{equation}
\begin{equation}
V_{2}f=\exp (\frac{1}{2\pi i}\alpha ^{k}\varphi _{kj}\alpha ^{j})f=\exp (%
\frac{1}{2\pi i}a^{k}\varphi _{kj}a^{j})f
\end{equation}
\begin{equation}
V_{3}f=f(A\alpha )
\end{equation}
\begin{equation}
V_{4}f=\exp (\frac{1}{2}b_{k}\theta ^{kj}b_{j})f
\end{equation}

The operators $V_{1}$ and $V_{4}$ relate $\widehat{\mu }$ and ch$(\widehat{E}%
),\mu $ and ch$(E)$, the operator $V_{2}$ relates ch$(\widehat{E})$ and ch$%
(E)$ and the operator $V_{3}$ should be included to take into account that
(31) is correct only for special identification of $L_{\theta }$ and $L_{%
\widehat{\theta }}$. It is easy to check that the operators $V_{k}$ can be
considered as linear canonical transformations. Recall that the operator $V$
acting on the Fock space is a linear canonical transformation if 
\begin{equation}
Va^{k}V^{-1}=A_{j}^{k}a^{j}+B^{kj}b_{j}
\end{equation}

\begin{equation}
Vb_{k}V^{-1}=C_{kj}a^{j}+D_{k}^{j}b_{j}
\end{equation}
or in matrix notations 
\begin{equation}
VaV^{-1}=Aa+Bb
\end{equation}
\begin{equation}
VbV^{-1}=Ca+Db
\end{equation}

It is easy to check that the $2n\times 2n$ matr\thinspace ices

\begin{equation}
\widetilde{V}=\left( 
\begin{array}{ll}
A & B \\ 
C & D
\end{array}
\right)
\end{equation}
belong to the group $O(n,n|\Bbb{C})$ i.e. they obey (11). The matrix $%
\widetilde{V}$ determines the operator $V$ up to a constant factor,
therefore the group of linear canonical transformations can be identified
with $O(n,n|\Bbb{C})\times \Bbb{C}^{*}$. It is easy to calculate that 
\begin{equation}
\widetilde{V}_{1}=\left( 
\begin{array}{ll}
1 & -\widehat{\theta } \\ 
0 & 1
\end{array}
\right) ,\widetilde{V}_{2}=\left( 
\begin{array}{ll}
1 & 0 \\ 
\Phi & 1
\end{array}
\right) ,\widetilde{V}_{3}=\left( 
\begin{array}{ll}
A & 0 \\ 
0 & (A^{t})^{-1}
\end{array}
\right) ,\widetilde{V}_{4}=\left( 
\begin{array}{ll}
1 & \theta \\ 
0 & 1
\end{array}
\right) .
\end{equation}

Here $\Phi $ stands for the matrix $(2\pi i)^{-1}\varphi _{kj}$. All these
matrices belong to $SO(n,n|\Bbb{R})$; hence their product belongs to the
same group. The operator $W=V_{1}V_{2}V_{3}V_{4}$ transforms the integral
lattice $\Lambda ^{ev}(D^{*})$ into itself; this means that $\widetilde{W}=%
\widetilde{V}_{1}\widetilde{V}_{2}\widetilde{V}_{3}\widetilde{V}_{4}$
belongs to $SO(n,n|\Bbb{Z})$. We see that

\begin{equation}
\widetilde{W}=\left( 
\begin{array}{ll}
A-\widehat{\theta }\Phi A & A\theta -\widehat{\theta }(\Phi A\theta
+(A^{t})^{-1}) \\ 
\Phi A & \Phi A\theta +(A^{t})^{-1}
\end{array}
\right) =\left( 
\begin{array}{ll}
M & N \\ 
R & S
\end{array}
\right)
\end{equation}
where $M,N,R,S$ are matrices with integral entries, 
\begin{equation}
M^{t}R+R^{t}M=N^{t}S+S^{t}N=0,\text{ }M^{t}S+R^{t}N=1.
\end{equation}

We conclude from (53)\ that 
\begin{equation}
\Phi A=R,A=\widehat{\theta }R+M,(A^{t})^{-1}=S-R\theta
\end{equation}

One can use these formulas to relate $\theta $ and $\widehat{\theta }:$%
\begin{equation}
(\widehat{\theta }R+\mu )\cdot (S^{t}+\theta R^{t})=1.
\end{equation}

Another relation between $\theta $ and $\widehat{\theta }$ is 
\begin{equation}
N=(M+\widehat{\theta }R)\theta -\widehat{\theta }S.
\end{equation}

Using (54) one can check that (56) and (57) are equivalent. We see that

\begin{equation}
\widehat{\theta }=(-M\theta +N)(R\theta -S)^{-1}
\end{equation}
and therefore $\widehat{\theta }=g\theta $ where $g\in SO(n,n|\Bbb{Z}).$ We
obtain that complete Morita equivalence of noncommutative tori is always of
the kind described in [3].

Let us consider a fibration having the space of irrational antisymmetric $%
n\times n$ matrices $\theta $ as a base and a group $K_{0}(T_{\theta })$ as
a fiber. It follows from the results we mentioned that the set of all
modules over $T_{\theta }$ with irrational $\theta $ is embedded in total
space of this fibration. Let us consider a curve in the base connecting $%
\theta $ with $g\theta $ where $g\in SO(n,n|\Bbb{Z})$ .It is easy to check
that the monodromy transformation corresponding to this curve connects
physically equivalent modules. The proof is based on a remark that for any
curve the monodromy can be considered as a linear canonical transformation
if the group $K_{0}(T_{\theta })$ is identified with a lattice in the Fock
space $\mathcal{F}^{*}$ .

\section{  Examples.}

Let us illustrate the above calculation on the example when $n=2k$ and $%
M=S=0,N=R=1.$ Then we get Morita equivalence of the tori $A_{\theta }$ and $%
A_{\theta ^{-1}}.$ The operator $W$ transforming $\mu $ into $\widehat{\mu }$
obeys $Wa^{i}W^{-1}=b_{i}$ and $Wb_{i}W^{-1}=a^{i}$ (i.e. it interchanges
creation and annihilation operators). If $\mu =\mu
(E)=\sum_{k}\sum_{i_{1}...i_{k}}\mu _{i_{1}...i_{k}}\alpha ^{i_{1}}...\alpha
^{i_{k}}$ then $\widehat{\mu }=\mu (\widehat{E})=W\mu =\sum_{l}\sum (*\mu
)_{j_{1}...j_{l}}\alpha ^{j_{1}}...\alpha ^{j_{l}}.$ In the case $n=2$ the
operator $W$ transforms $\mu =p+q\alpha ^{1}\alpha ^{2}$ into $\widehat{\mu }%
=-q-p\alpha ^{1}\alpha ^{2}.$ Corresponding Chern characters are 
\begin{equation}
\limfunc{ch}=e^{\theta b_{1}b_{2}}\mu =(p-q\theta )+q\alpha ^{1}\alpha ^{2}
\end{equation}
\begin{equation}
\widehat{\limfunc{ch}}=e^{\theta ^{-1}b_{1}b_{2}}\widehat{\mu }=(-q+p\theta
^{-1})+p\alpha ^{1}\alpha ^{2}.
\end{equation}

We see that dimensions of modules $E$ and $\widehat{E}$ are equal to $\dim
E=p-q\theta $ and $\dim \widehat{E}=-q+p\theta ^{-1}=\theta ^{-1}\cdot \dim
E $ correspondingly.

Every projective module over two-dimensional noncommutative torus can be
equipped with a constant curvature connection. One can calculate its
curvature $F$ using (37) and (58). We get 
\begin{equation}
F=2\pi i\frac{q}{p-q\theta }\alpha ^{1}\alpha ^{2}
\end{equation}

\begin{equation}
\widehat{F}=2\pi i\frac{p\theta }{p-q\theta }\alpha ^{1}\alpha ^{2}
\end{equation}
for modules $E$ and $\widehat{E}$ respectively. Using (55) we see that $%
A=\theta ^{-1},$ $\Phi =\theta .$ Hence if $F^{^{\prime }}$ is obtained from 
$F$ by means of the change of variables $\widetilde{\alpha }=A\cdot \alpha
=\theta ^{-1}\cdot \alpha $ we should have $F^{^{\prime }}+2\pi i\cdot
\theta =\widehat{F}.$ This is true: 
\begin{equation}
\frac{q}{p-q\theta }\theta ^{2}+\theta =\frac{p\theta }{p-q\theta }
\end{equation}

In the case $n=4$ the operator $W$ transforms $\mu =p+\frac{1}{2}%
q_{ij}\alpha ^{i}\alpha ^{j}+s\alpha ^{1}\alpha ^{2}\alpha ^{3}\alpha ^{4}$
into $_{{}}\widehat{\mu }=s+\frac{1}{2}(*q)_{ij}\alpha ^{i}\alpha
^{j}+p\alpha ^{1}\alpha ^{2}\alpha ^{3}\alpha ^{4}.$ Corresponding Chern
characters are

\begin{equation}
\limfunc{ch}=p+\frac{1}{2}\limfunc{tr}\theta q+s\underline{\theta }+\frac{1}{%
2}(q+*\theta s)_{ij}\alpha ^{i}\alpha ^{j}+s\alpha ^{1}\alpha ^{2}\alpha
^{3}\alpha ^{4}
\end{equation}
\begin{equation}
\widehat{\limfunc{ch}}=s+\frac{1}{2}\limfunc{tr}\theta ^{-1}(*q)+p\underline{%
(\theta ^{-1})}+\frac{1}{2}(*q+*(\theta ^{-1})p)_{ij}\alpha ^{i}\alpha
^{j}+p\alpha ^{1}\alpha ^{2}\alpha ^{3}\alpha ^{4}
\end{equation}

Here \underline{$\theta $} stands for the Pfaffian of $\theta $ (i.e. 
\underline{$\theta $}$^{2}=\det \theta _{ij}),\underline{(\theta ^{-1})}=(%
\underline{\theta )}^{-1}.$ We obtain from (64),(65) that $\dim \widehat{E}%
=(\dim E)\cdot \underline{\theta }^{-1}.$ If there exists a constant
curvature connection in $E$, then there exists such a connection in $%
\widehat{E}$ and corresponding curvatures up to a factor $2\pi i$ can be
identified with matrices 
\begin{equation}
F=\frac{q+(*\theta )s}{\dim E},\widehat{F}=\frac{*q+*(\theta ^{-1})p}{\dim 
\widehat{E}}
\end{equation}

Taking into account that $A=\theta ^{-1},\Phi =\theta $ we can represent the
relation (31) in the form 
\begin{equation}
-\theta F\theta +\theta =\widehat{F}
\end{equation}

One can check (67) directly using a non-trivial identify 
\begin{equation}
-\theta q\theta +\frac{1}{2}\theta \limfunc{Tr}\theta q=*q\underline{\theta }
\end{equation}
that is valid for all antisymmetric $4\times 4$ matrices $\theta ,q.$

\section{ Relation to BFSS\ M(atrix)\ model.}

Let us study more thoroughly the action functional (27) in the case when
there exists a constant curvature connection $\nabla _{\alpha }$ in $A$%
-module $E$. Then any other connection can be represented as $\nabla
_{\alpha }+X_{\alpha }$ where $X_{\alpha }\in $ End$_{A}E$ and its curvature 
$F_{\alpha \beta }$ can be written in the form $F_{\alpha \beta }=f_{\alpha
\beta }\cdot 1+\widetilde{\delta }_{\alpha }X_{\beta }-\widetilde{\delta }%
_{\beta }X_{\alpha }+[X_{\alpha },X_{\beta }].$ where $f_{\alpha \beta }$
stands for the curvature of $\nabla _{\alpha }$ and $\widetilde{\delta }%
_{\alpha }X$ is defined as $[\nabla _{\alpha },X].$ We can consider (27) as
a functional depending on $X_{\alpha }\in $ End$_{A}E,\alpha =1,...,10,$ and 
$\Psi ^{i}\in \Pi $End$_{A}E,i=1,...,16;$ it is equal to 
\begin{equation}
J=\sum \limfunc{Tr}(\widetilde{\delta }_{\alpha }X_{\beta }-\widetilde{%
\delta }_{\beta }X_{\alpha }+[X_{\alpha },X_{\beta }]+(f_{\alpha \beta
}+\varphi _{\alpha \beta })\cdot 1)^{2}+2\sum \limfunc{Tr}\Psi ^{i}\Gamma
_{ij}^{\alpha }(\widetilde{\delta }_{\alpha }\Psi ^{j}+[X_{\alpha },\Psi
^{j}]).
\end{equation}

We used orthonormal coordinate system in $L$ in the expression for $J.$ It
is easy to rewrite this expression in arbitrary coordinate system by means
of relations $X_{\alpha }=e_{\alpha }^{a}X_{a},X_{a}=e_{a}^{\alpha
}X_{\alpha },\widetilde{\delta }_{\alpha }=e_{\alpha }^{a}\widetilde{\delta }%
_{a,}\widetilde{\delta }_{a}=e_{a}^{\alpha }\widetilde{\delta }_{\alpha }.$
In the case when End$_{A}E$ can be identified with $n$-dimensional
noncommutative torus $T_{\rho }$ we can consider elements of End$_{A}E$ as
functions on (commutative) torus $L_{\rho }^{*}/D^{*};$ then for appropriate
choice of coordinate systems on $L$ and on $L_{\rho }$ we can identify $%
\widetilde{\delta }_{a}$ for $1\leq a\leq n$ with partial derivatives $\frac{%
\partial }{\partial \sigma ^{a}}$ and assume that $\widetilde{\delta }_{a}=0$
for $a>n$. (We denote coordinates on $L_{\rho }^{*}/D^{*}$ by $\sigma
^{1},...,\sigma ^{n}$.) The same is correct if End$_{A}E$ is an algebra of $%
N\times N$ matrices with entries from $T_{\rho }.$ Similar result can be
proved also in general case . ( It follows from well known results that for
modules considered in Sec. 3 End$_{A}E$ can be represented as an algebra of
functions on finitely generated abelian group with multiplication rule (15)
where $\vartheta _{XY}$ is a bicharacter.) Let us write down the expression
for the functional $J$ in the latter case restricting ourselves to the
bosonic part of $J$ for simplicity. 
\begin{equation}
J_{bos}=\sum_{a,b}\limfunc{Tr}F_{ab}F^{ab}
\end{equation}
\begin{equation}
F_{ab}=\frac{\partial }{\partial \sigma ^{a}}X_{b}-\frac{\partial }{\partial
\sigma ^{b}}X_{a}+[X_{a},X_{b}]+(f_{ab}+\varphi _{ab})\cdot 1\limfunc{for}%
1\leq a,b\leq n,
\end{equation}
\begin{equation}
F_{ab}=\frac{\partial }{\partial \sigma ^{a}}X_{b}+[X_{a},X_{b}]+(f_{ab}+%
\varphi _{ab})\cdot 1\limfunc{for}1\leq a\leq n,b>n,
\end{equation}
\begin{equation}
F_{ab}=[X_{a},X_{b}]+(f_{ab}+\varphi _{ab})\cdot 1\limfunc{for}a>n,b>n.
\end{equation}

Tr stands for composition of matrix trace tr and integration over $\sigma
^{1},...,\sigma ^{n}$ (i.e. $\limfunc{Tr}R=\int_{L_{\rho }/D}\limfunc{tr}%
R(\sigma ^{1},...,\sigma ^{n})d^{n}\sigma $). In (70)\ we raise indices by
means of metric tensor $g_{ab}=e_{a}^{\alpha }e_{b}^{\alpha }$. We can
replace this Euclidean metric tensor with metric tensor having Lorentzian
signature (i.e. perform ''Wick rotation '') . This corresponds to
consideraton of compactification of BFSS\ M(atrix) model.

We have proven that Morita equivalence gives us equivalence between toroidal
compactifications of IKKT\ M(atrix) model. It is clear that performing
corresponding Wick rotations we obtain equivalent compactifications of BFSS
model. We will analyze this equivalence thoroughly in a forthcoming paper
[10].

\centerline \textbf{APPENDIX}

Let us start with the following general observation about supermatrix
models. We consider an action functional of the form 
\[
S=\sum a_{\alpha _1...\alpha _n} Tr X^{\alpha _1}...X^{\alpha _n}. 
\]
defined on a set of matrices $X^{\alpha}$ , $1\leq \alpha \leq c$. We
consider $X^{\alpha}$ as matrices of (even or odd) Hermitian operators
acting in the space $\mathbf{C}^{K|L}$; in other words they are $(K|L)\times
(K|L)$-matrices. Corresponding partition function $Z$ is defined by means of
integration of $\exp (-S)$ over all Hermitian matrices. Then it is easy to
check that $Z$ depends only on the difference $N=K-L$. One of the ways to
prove this fact is to apply the standard representation of $Z$ in terms of a
sum over all ribbon graphs. The size of the matrix enters in this
representation only as $Tr 1=K-L$.

Now we can consider toroidal compactifications of IKKT or BFSS matrix models
taking into account the above remark. It was shown in [1] that such
compactifications are related to connections in projective modules over
commutative or non-commutative torus. More precisely, the equations arising
in compactification problem cannot be solved by means of finite-dimensional
matrices, but can be solved in terms of Hermitian operators in Hilbert
spaces with action of (noncommutative) torus. These infinite-dimensional
solutions can be used to obtain approximate finite-dimensional solutions.
Now we see that we can work with the same success in $\mathbf{Z}_{2}$-graded
Hilbert spaces. Then the approximate finite-dimensional solutions are
supermatrices, but we have seen that the matrix model based on $(K|L)\times
(K|L)$ supermatrices is equivalent to theory based on $(K-L)\times (K-L)$
matrices.

Let us remark now that the theory of $\mathbf{Z}_{2}$-graded projective
modules is closely related to $K$-theory. The group $K_{0}(A)$ where $A$ is
an associative algebra, can be define as a group of $\mathbf{Z}_{2}$-graded
projective modules with equivalence relation $E\oplus \Pi E\backsim 0$.
(Here $\Pi $ stand for parity reversion. The operation in $K_{0}(A)$ is
defined as direct sum; the equivalence relation $E\oplus \Pi E\backsim 0$
permits us to identify element $-E$ with $\Pi E$.) One can relate $K$-theory
also to the theory of differential $\mathbf{Z}_{2}$-graded projective
modules, i.e.$\mathbf{Z}_{2}$-graded modules equipped with an odd ( parity
reversing) $A$ -linear map $Q,$ obeying $Q^{2}=0$ .We can say that a $%
\mathbf{Z}_{2}$-graded module $\mathcal{E}$ determines zero element of the
group $K_{0}(A)$ iff one can introduce a parity reversing differential $Q$
on $\mathcal{E}$ in such a way that

\begin{equation}
\limfunc{Im}Q=\limfunc{Ker}Q
\end{equation}
(i.e. $Q$ has trivial homology) and $\limfunc{Im}Q$ is a direct summand in $%
\mathcal{E}$. (In this case we can write that $\mathcal{E}=E\oplus \limfunc{%
Im}Q$; the map $Q$ identifies $\limfunc{Im}Q$ and $\Pi E$.)

\centerline 
\textbf{Acknowledgements. }

I am deeply indebted to M. Rieffel for numerous consultations on
multidimensional noncommutative tori that were thoroughly analyzed in his
papers. I am grateful to A. Astashkevich and N. Nekrasov for interesting
discussions. I was supported in part by NSF grant DMS 95-00704 and ( during
my visit to the Institute for Theoretical Physics, Santa Barbara) by NSF
grant PHY94-07194.

\textbf{REFERENCES. }

1. Connes, A., Douglas, M., and Schwarz, A., \textbf{Noncommutative Geometry
and Matrix Theory: Compactification on Tori}, hep-th/9711162. Published in
JHEP electronic journal.

2. Douglas, M. and Hull, C., \textbf{$D$-branes and non-commutative geometry}%
, hep-th/9711165.

Kawano, T. and Okuyama, K., \textbf{Matrix Theory on Noncommutative Torus},
hep-th/9803044.

Cheu, Y.-K. E. and Krogh, M., \textbf{Noncommutative Geometry from 0-branes
in a Background B-field}, hep-th/9803031.

Berkooz, M., \textbf{Non-local Field Theories and the Non-commutative Torus}%
, hep-th/980206.

Li, M., \textbf{Comments on Supersymmetric Yang-Mills Theory on a
Noncommutative Torus}, hep-th/9802052.

Casalbuoni, R., \textbf{Algebraic treatment of compactification on
noncommutative tori}, hep-th/9801170.

Ho, P.-M. and Wu, Y.-S., \textbf{Noncommutative Gauge Theories in Matrix
Theory}, hep-th/9801147.

Krogh, M., \textbf{A Matrix Model for Heterotic $Spin(32)/Z_{2}$ and Type I
String Theory}, hep-th/9801034.

Ho, P.-M., Wu, Y.-Y., and Wu, Y.-S. , \textbf{Towards a Noncommutative
Geometric Approach to Matrix Compactification}, hep-th/9712201.

Leigh, R. G. and Rozali, M., \textbf{A Note on Six-Dimensional Gauge Theories%
}, hep-th/9712168

Aharony, O., Berkooz, M., and Seiberg, N., \textbf{Light-Cone Description of
(2,0) Superconformal Theories in Six Dimensions, }hep-th/9712117.\textbf{\ }

Obers, N. A., Pioline, B., and Rabinovici, E.,\textbf{\ M-Theory and
U-duality on $T^{d}$ with Gauge Backgrounds, }hep-th/9712084.

Hull, C. M.,\textbf{\ U-Duality and BPS Spectrum of Super Yang-Mills Theory
and M-Theory, }hep-th/9712075.

\smallskip Blau, M. and O'Laughlin, M., \textbf{Aspects of U-Duality in
Matrix Theory, }hep-th/9712047.

3. Rieffel, M. and Schwarz, A.,\textbf{\ Morita equivalence of
multidimensional noncommutative tori,} q-alg/ 9803057.

4.Nekrasov, N. and Schwarz, A.,\textbf{\ Instantons on noncommutative $R^{4}$%
, and (2,0) superconformal six dimensional theory, }hep-th/9802068.

5. Astashkevich, A., Nekrasov, N., and Schwarz, A., in preparation.

6. Connes, A.,\textbf{\ Noncommutative geometry, }Academic Press, 1994.

\ 7. Connes, A. and Rieffel, M.\textbf{\ Yang-Mills for noncommutative
two-tori, }in : Operator algebras and Mathematical Physics ( Iova City,
Iova,1985), pp. 237-266, Contemp. Math. Oper. Algebra. Math.Phys. 62, AMS
1987.\textbf{\ }

Rieffel, M. A.,\textbf{\ Induced representations of $C^{*}$-algebras, }%
Advances Math. 13 (1974), 176--257.

Rieffel, M. A.\textbf{, $C^{*}$-algebras associated with irrational
rotations,} Pacific J. Math. 93 (1981), 415--429.

Rieffel, M. A.,\textbf{\ Morita equivalence for operator algebras,} in
``Operator Algebras and Applications'' (R. V. Kadison, ed.) Proc. Symp. Pure
Math. 38, Amer. Math. Soc., Providence, 1982.

Rieffel, M. A.,\textbf{\ Projective modules over higher-dimensional
non-commutative tori,} Canadian J. Math. 40 (1988), 257--338.

Rieffel, M. A.,\textbf{\ Non-commutative tori --- a case study of
non-commutative differentiable manifolds, }Contemporary Math. 105 (1990),
191--211.

Elliott, G. A.,\textbf{\ On the K-theory of the $C^{*}$-algebras generated
by a projective representation of a torsion-free discrete abelian group, }in
``Operator Algebras and Group Representations'' 157--184, Pitman, London,
1984.

8. Banks, T., Fishler, W., Shenker, S., and Susskind, L., Phys. Rev. D55
(1997) 5112-5128, hep-th /9610043

9. Ishibashi, N.Kawai, H., Kitazava, I., and Tsuchiya, A., Nucl. Phys. B492
(1997) 467-491

10.Konechny, A and Schwarz, A., in preparation.

\end{document}